\def\MPL #1 #2 #3 {Mod.~Phys.~Lett.~{\bf#1},\  #2 (#3)}
\def\NPB #1 #2 #3 {Nucl.~Phys.~{\bf#1},\  #2 (#3)}
\def\PLB #1 #2 #3 {Phys.~Lett.~{\bf#1},\  #2 (#3)}
\def\PR #1 #2 #3 {Phys.~Rep.~{\bf#1},\ #2 (#3)}
\def\PRD #1 #2 #3 {Phys.~Rev.~{\bf#1},\  #2 (#3)}
\def\PRL #1 #2 #3 {Phys.~Rev.~Lett.~{\bf#1},\  #2 (#3)}
\def\RMP #1 #2 #3 {Rev.~Mod.~Phys.~{\bf#1},\  #2 (#3)}
\def\ZP #1 #2 #3 {Z.~Phys.~{\bf#1},\  #2 (#3)}
\def\IJMP #1 #2 #3 {Int.~J.~Mod.~Phys.~{\bf#1},\  #2 (#3)}
\def\cmd{(c^2-d^2)}
\def\ocp{{\cal O}_{CP}}

\def\cpd{(c^2+d^2)}
\def\asym{\alpha}
\def\bsym{\beta} 
\def\asyms{\asym_S}
\def\asymb{\asym_B}
\def\bsyms{\bsym_S}
\def\bsymb{\bsym_B}

\def\wstar{W^{\star}}
\def\gamv{\gamma_5}

\def\h{h}
\def\mh{m_{\h}}

\def\lsim{\mathrel{\raise.3ex\hbox{$<$\kern-.75em\lower1ex\hbox{$\sim$}}}}
\def\gsim{\mathrel{\raise.3ex\hbox{$>$\kern-.75em\lower1ex\hbox{$\sim$}}}}
\def\@versim#1#2{\vcenter{\offinterlineskip
        \ialign{$\m@th#1\hfil##\hfil$\crcr#2\crcr\sim\crcr } }}

\def\wstar{W^\star}

\def\ie{{\it i.e.}}

\def\gam{\gamma}

\def\anti{\overline}

\def\fbi{~{\rm fb}^{-1}}

\def\gev{\,{\rm GeV}}

\def\hsm{\phi^0}

\def\hl{h^0}

\def\ha{A^0}

\def\mha{m_{\ha}}

\def\tanb{\tan\beta}

\def\mt{m_t}

\def\wp{W^+}
\def\wm{W^-}

\def\h{h}
\def\mh{m_{\h}}

\documentstyle[12pt,equations]{article}
\def\ie{{\it i.e.}}

\def\9{\phantom 0}      %%% for lining up numbers in columns
\renewcommand\linebreak{\unskip\break} %% breaks line & still justifies
\begin{document}
\newlength{\captsize} \let\captsize=\small % use \let\normalsize=\captsize
\newlength{\captwidth}                     % just before \caption{ ...
%%%%%%%%%%%%%%%%%%%%%%%%%%%%%%%%%%%%%%%%%%%%%%%%%%%%%%%%%%%%%%

%\preprint{
%
\font\fortssbx=cmssbx10 scaled \magstep2
\hbox to \hsize{
%
%\special{psfile=uwlogo.ps
% hscale=8000 vscale=8000
% hoffset=-12 voffset=-2}
%\hskip.5in \raise.1in
%
$\vcenter{
\hbox{\fortssbx University of California - Davis}
%\hbox{\fortssbx University of Wisconsin - Madison}
}$
\hfill
$\vcenter{
\hbox{\bf UCD-96-5} 
%\hbox{\bf MADPH-95-884} 
%\hbox{\bf IUHET-299}
\hbox{February, 1996}
}$
}
%}

%
\medskip
\begin{center}
\bf
DETERMINING THE CP NATURE OF A NEUTRAL HIGGS BOSON AT THE LHC
\\
\rm
\vskip1pc
{\bf John F. Gunion and Xiao-Gang He}\\
\medskip
{\em Davis Institute for High Energy Physics}\\
{\em Department of Physics, University of California, Davis, CA 95616}\\
\end{center}

\begin{abstract}
We demonstrate that certain weighted moments of
$t\anti t \h$ production can provide
a determination of the CP nature of a light neutral Higgs
boson produced and detected in this mode at the CERN LHC.
\end{abstract}

\section{Introduction}

\indent\indent 
It is well-known 
\cite{hhg,dpfreport} that detection of a Higgs boson $\h$ with couplings 
roughly like those of the Standard Model (SM) Higgs ($\hsm$)
will be possible at the LHC, the detection mode depending
upon the mass $\mh$. If $\mh\lsim 130\gev$,
the primary techniques for observing such an $\h$ rely
on $gg\to\h$, $\wstar\to W\h$,
and $gg\to t\anti t\h$, with $\h\to \gam\gam$ and, 
for the latter two modes, $\h\to b\anti b$. If the $\h$ is part
of a larger (non-SM) Higgs sector,
deviations of the $W\h$ and $t\anti t \h$ production rates
from SM expectations will be present, but difficult to interpret. 
A direct determination of whether the $\h$ is CP-even (as predicted
for the SM $\hsm$ and the minimal supersymmetric model $\hl$) 
would be especially crucial to unravelling the situation.
The only procedure proposed to date
for directly determining the CP nature of a light $\h$
employs proton beam polarization asymmetries that 
are only sufficiently large if proton polarization implies substantial 
polarization for the colliding gluons
in the $gg\to \h$ process \cite{ggy}. Here, we demonstrate
that certain weighted moments of the cross section for $t\anti t\h$
production are sensitive to the relative magnitudes
of the CP-even and CP-odd coupling 
coefficients $c$ and $d$ (respectively) in the $t\anti t\h$
interaction Lagrangian, ${\cal L}\equiv \anti t (c+id\gamv) t \h$. The 
accuracy with which these moments can be measured experimentally
at the LHC (assuming an accumulated, detector-summed
luminosity of $L=600\fbi$) is sufficient that 
the CP-even SM Higgs boson can be shown to be inconsistent
with an equal mixture of CP-odd and CP-even components at 
the $\sim 1.5\sigma-2\sigma$ statistical level, and inconsistent with
a purely CP-odd state at the $\gsim 7\sigma$ statistical level.

\section{The Technique}

\indent\indent 
Consider a general quark-antiquark-Higgs
coupling of the form $\anti Q (c+id\gamv) Q \h$,
where $c$ and $d$ are both taken to be real; $c$ and $d$
determine the CP-even and CP-odd components of the coupling, respectively.
By explicit calculation, it can be demonstrated that the spin-averaged
cross sections for $ gg\to Q\anti Q\h$ and $q\anti q\to Q\anti Q\h$
production contain no terms proportional
to $cd$. However, they do contain terms proportional
to both $\cpd$ and $\cmd$. Since the $\cmd$ 
terms are multiplied by $m_q^2$,\footnote{This is in addition
to the $m_q^2$ factor contained in the $c^2$ and $d^2$ squared 
couplings themselves.}
sensitivity to $\cmd$ 
terms is only significant if $m_q^2$ is of the same order as the other
invariant masses-squared for the subprocess.  The latter
are fairly large, being set by the scale $\mh^2$, implying that only 
$t\anti t \h$ production will have substantial sensitivity to $\cmd$. 
For convenience, we 
adopt the conventional normalization of $\cpd=1$ in our tabulations below.

In order to isolate the $\cmd$ terms in the production
amplitude-squared in a manner that is free of systematic
uncertainties associated with the overall production rate,
we compute the ratio:
\begin{equation}
\asym[\ocp]\equiv {\int [\ocp]\left\{d\sigma(pp\to t\anti t X)/
dPS\right\}dPS\over  \int\left\{d\sigma(pp\to t\anti t X)/
dPS\right\}dPS}\,,
\label{adef}
\end{equation}
where $\ocp$ is an operator designed to maximize sensitivity
of $\asym$ to the $\cmd$ term. We have found that
a variety of simple operators offer substantial sensitivity
to $\cmd$. The best of those that we have examined are 
\begin{equation}
\begin{array}{llll}
a_1 = {(\vec p_t\times \widehat n)\cdot (\vec p_{\bar t} \times \widehat n)
\over |(\vec p_t\times \widehat n)\cdot
(\vec p_{\bar t}\times \widehat n)|}&
b_1 = {(\vec p_t\times \widehat n)\cdot (\vec p_{\bar t} \times \widehat n)
\over p_t^T p_{\bar t}^T } &
b_2 = {(\vec p_t\times \widehat n)\cdot (\vec p_{\bar t} \times \widehat n)
\over |\vec p_t| |\vec p_{\bar t}|} \\
a_2= { p_t^{\, x}  p_{\bar t}^{\, x}\over 
| p_t^{\, x}  p_{\bar t}^{\, x}|}&
b_3 = { p_t^{\,x}  p_{\bar t}^{\,x}\over p_t^T p_{\bar t}^T }&
b_4 = { p_t^{\,z}  p_{\bar t}^{\,z}\over |\vec p_t|
|\vec p_{\bar t}|} 
\label{ocpdefs}
\end{array}
\end{equation}
where $p_{t,\bar t}^T$ denote the magnitudes of the $t$ and $\anti t$
transverse momenta. In Eq.~\ref{ocpdefs}, $\widehat n$ is a unit vector
in the direction of the beam line and defines the $z$ axis. One can
also employ the $y$-component analogues of $a_2$ and $b_3$.
Since all these operators will have
somewhat different systematic uncertainties, 
it will be useful to analyze the CP properties using all of them
(and, perhaps, others as well). 
Higher moments of the operators were also considered, but
in all cases the linear moments were the most sensitive to
changes in the $c$ vs. $d$ weighting. 

The critical question with regard to the usefulness of
a given $\asym$ is the accuracy with which it can be measured
relative to the predicted changes in $\asym$ as a function of the
CP nature of the $\h$. In general, there will
be background as well as $t\anti t\h$ signal contributions
in any $t\anti t X$ channel ($X=\gam\gam$ or $b\anti b$).
We define $\asyms$ and $\asymb$ to be the value of $\asym$
as defined in Eq.~\ref{adef} for the signal and background
cross sections on their own. We also define
\begin{equation}
\bsym[\ocp]\equiv {\int [\ocp]^2\left\{d\sigma(pp\to t\anti t X)/
dPS\right\}dPS\over  \int\left\{d\sigma(pp\to t\anti t X)/
dPS\right\}dPS}\,, 
\label{bdef}
\end{equation}
and the $\bsyms$ and $\bsymb$ values of $\bsym$ for the signal
and background individually.  Then, one finds that the experimental
error for $\asyms$ is given by
\begin{equation}
\delta\asyms= S^{-1/2}\left[\bsyms-\asyms^2+{B\over S}
\left(\bsymb-2\asymb\asyms+\asyms^2\right)\right]^{1/2}\,,
\label{dasymdef}
\end{equation}
where $S$ and $B$ are the total number of signal and background
events, respectively, in the $t\anti t X$ channel being considered. 
This result assumes that $B$, $\asymb$ and $\bsymb$ can be experimentally
measured (using data with $M_X$ not in the vicinity of $\mh$) and/or
calculated with high precision.
Note that $\bsym=1$ for the $a_i$ operators.

\begin{table}[hbt]
\caption[fake]{$t\anti t\gam\gam$ channel 
$\asymb$ and $\bsymb$ values, assuming $\mh=100\gev$.}
\begin{center}
\begin{tabular}{|c|cccccc|}
\hline
$\ocp$ & $a_1$ & $b_1$ & $b_2$ & $a_2$ & $b_3$ & $b_4$ \\
\hline
$\asymb$ & $-0.863$ & $-0.796$ & $-0.249$ & $-0.698$ & $-0.404$ & $0.130$ \\
$\bsymb$ & 1 & 0.806 & 0.127 & 1 & 0.332 & 0.411 \\
\hline
\end{tabular}
\end{center}
\label{abbkgndvaluesgamgam}
\end{table}

\begin{table}[hbt]
\caption[fake]{Values of $\asyms$ and $\bsyms$ for the 
$t\anti t\h$ production, assuming $\mh=100\gev$,
for the cases: $c=1,d=0$; $c=d=1/\sqrt2$; $c=0,d=1$. Also given
is $S^{1/2}\delta\asyms$ in the limit $B=0$.}
\begin{center}
\begin{tabular}{|l|rrrrrr|}
\hline
$\ocp$ & $a_1$ & $b_1$ & $b_2$ & $a_2$ & $b_3$ & $b_4$ \\
\hline
$c=1,d=0$ & & & & & & \\
$\asyms$ & $-0.810$ & $-0.718$ & $-0.269$ & $-0.619$ & $-0.359$ & $0.292$ \\
$\bsyms$ & 1 & $0.736$ & $0.151$ & 1 & $0.308$ & $0.376$ \\
$S^{1/2}\delta\asyms$ & 0.586 & 0.469 & 0.280 & 0.785 & 0.424 & 0.539 \\
\hline
$c=d=1/\sqrt 2$ & & & & & & \\
$\asyms$ & $-0.742$ & $-0.654$ & $-0.243$ & $-0.562$ & $-0.327$ & 0.228 \\
$\bsyms$ & 1 & $0.707$ & $0.139$ & 1 & $0.302$ & $0.372$ \\
$S^{1/2}\delta\asyms$ & 0.671 & 0.528 & 0.283 & 0.827 & 0.442 & 0.566 \\
\hline
$c=0,d=1$ & & & & & & \\
$\asyms$ & $-0.486$ & $-0.407$ & $-0.147$ & $-0.335$ & $-0.200$ & $-0.005$ \\
$\bsyms$ & 1 & 0.593 & $0.096$ & 1 & 0.272 & 0.371 \\
$S^{1/2}\delta\asyms$ & 0.874 & 0.654 & 0.272 & 0.942 & 0.482 & 0.609 \\
\hline
\end{tabular}
\end{center}
\label{absigvalues}
\end{table}

Let us now focus on the $X=\gam\gam$ channel. Detection
of a Higgs boson in the $t\anti t \gam\gam$ final state was originally proposed
in Ref.~\cite{GMP} and has been thoroughly studied by both
the ATLAS and CMS collaborations in their Technical Proposals \cite{ATLAS,CMS}.
The final state employed is that in which one $t$ decays semi-leptonically,
while the other decays hadronically. This allows identification
of the $t$ vs. the $\anti t$, and reconstruction
of the transverse momenta of both the $t$ and $\anti t$. This is already
sufficient for defining $a_{1,2}$ and $b_{1,3}$ in Eq.~\ref{ocpdefs}. 
The $b_{2,4}$
operators of Eq.~\ref{ocpdefs} require knowledge of the full $t$ and $\anti t$
three momenta. For the hadronically decaying $t$, the three jets
can be used for this determination. Determination of
the $z$ component of momentum of the leptonically decaying $t$ is
subject to the usual two-fold ambiguity in determining the $z$ component
of the momentum of the unobserved neutrino using the missing transverse
energy, $m_\nu=0$, and the known value of $\mt$. The algorithm in which
$p_\nu^z$ is chosen so as to minimize the overall rapidity of the $t\anti t\h$
system yields the correct solution a large fraction of the time.

We first present the values of $\asym$ and $\bsym$ for signal and background,
for the various different $\ocp$ operators listed in Eq.~\ref{ocpdefs},
assuming a Higgs boson mass of $\mh=100\gev$.
These values include mild cuts on the outgoing $t$, $\anti t$
and photons of: $|y_{t,\bar t,\gam}|<4$, $p^T_{\gam}>\mh/4$.
In actual practice, the detector collaborations must compute
the expected numbers using their full cuts and resolutions.
Results for $\asymb$ and $\bsymb$ are given in
Table~\ref{abbkgndvaluesgamgam}, and results for $\asyms$ and $\bsyms$
are given in Table~\ref{absigvalues} for several choices
of $c,d$. Also given are the values of $S^{1/2}\delta\asyms$ 
in the limit of zero background.  These latter values
give a first indication of the relative sensitivity of the various
$\ocp$ operators to different values of $c,d$.

\begin{table}[hbt]
\caption[fake]{$t\anti t\gam\gam$ channel, $\mh=100\gev$: 
discrimination powers in the limit $B=0$.}
\begin{center}
\begin{tabular}{|l|rrrrrr|}
\hline
$\ocp$ & $a_1$ & $b_1$ & $b_2$ & $a_2$ & $b_3$ & $b_4$ \\
\hline
%results for average error
%$D_1/\sqrt S$ & 0.109 & 0.129 & 0.091 & 0.071 & 0.074 & 0.115 \\
%$D_2/\sqrt S$ & 0.445 & 0.554 & 0.442 & 0.329 & 0.350 & 0.517 \\
%results for SM error 
$D_1/\sqrt S$ & 0.117 & 0.137 & 0.091 & 0.073 & 0.076 & 0.118 \\
$D_2/\sqrt S$ & 0.554 & 0.663 & 0.436 & 0.362 & 0.374 & 0.551 \\
\hline
\end{tabular}
\end{center}
\label{discpower}
\end{table}

In order to better interpret these results, we define the two discrimination
powers:
\begin{eqnarray}
D_1&\equiv &{|\asyms(c=1,d=0)-\asyms(c=d=1/\sqrt 2)|\over 
%{1\over 2} [\delta\asyms(c=1,d=0)+\delta\asyms(c=d=1/\sqrt 2)]}\,, \nonumber \\
\delta\asyms(c=1,d=0)}\,, \nonumber \\
D_2&\equiv &{|\asyms(c=1,d=0)-\asyms(c=0,d=1)|\over 
%{1\over 2}[\delta\asyms(c=1,d=0)+\delta\asyms(c=0,d=1)]}\,,
\delta\asyms(c=1,d=0)}\,, \nonumber \\
\label{ddefs}
\end{eqnarray}
which indicate the ability of these observables to separate the
SM-like case of $c=1,d=0$ from the equal mixture case, $c=d=1/\sqrt 2$,
and from the pure CP-odd case of $c=0,d=1$. These 
definitions of $D_{1,2}$ are those appropriate if the observed 
Higgs boson is CP-even (so that it is appropriate to use
$\delta\asyms(c=1,d=0)$ in computing the experimental error).\footnote{If the 
$\h$ is pure CP-odd, $D_2$ should be defined using $\delta\asyms(c=0,d=1)$;
if it is an equal CP-even/CP-odd mixture, $D_1$ should be defined using
$\delta\asyms(c=d=1/\sqrt 2)$.}
Values for $D_{1,2}/\sqrt S$ are tabulated
in Table~\ref{discpower} in the limit where $B=0$ (no background). 
From this table, it is apparent that
the operators $a_1$, $b_1$, $b_2$ and $b_4$ will be most useful,
$b_1$ probably being the best, both in that it has the largest
discrimination power, and also in that it can be constructed
without use of $z$-component momenta.

\begin{table}[hbt]
\caption[fake]{$t\anti t\gam\gam$ channel, $\mh=100\gev$: 
the discrimination powers $D_{1,2}$ for $S=130$ and $B=21$.}
\begin{center}
\begin{tabular}{|l|rrrrrr|}
\hline
$\ocp$ & $a_1$ & $b_1$ & $b_2$ & $a_2$ & $b_3$ & $b_4$ \\
\hline
%results for average errors
%$D_1$ & 1.18 & 1.39 & 0.97 & 0.76 & 0.79 & 1.19 \\
%$D_2$ & 4.84 & 5.95 & 4.71 & 3.54 & 3.72 & 5.37 \\
% results for SM case
$D_1$ & 1.26 & 1.47 & 0.97 & 0.78 & 0.81 & 1.21 \\
$D_2$ & 5.97 & 7.10 & 4.67 & 3.87 & 3.97 & 5.65 \\
\hline
\end{tabular}
\end{center}
\label{disctrue}
\end{table}

Of course, in practice we must include the background in order
to obtain the true value of $\delta\asyms$, see Eq.~\ref{dasymdef}.
Let us focus on the case where
the $\h$ has SM-like couplings and ask how well we can measure
$\asyms$ in this case.  For this purpose, we employ the results
from the CMS experimental Technical Proposal, Figure 12.8
(in which $L=162.5\fbi$ is assumed),
scaled to the current standard benchmark in which it is assumed
that each detector (ATLAS and CMS) will separately accumulate an
integrated luminosity of $L=300\fbi$, corresponding to a total detector-summed
luminosity of $L=600\fbi$. For our estimates we will sum events
in a $5\gev$ interval centered at $M_{\gam\gam}=\mh=100\gev$. From
the above-referenced Fig. 12.8, one obtains, after rescaling,
$S=259$ and $B=42$
for the sum of the $W\gam\gam$ and $t\anti t \gam\gam$ modes.  At
the LHC, these two modes contribute almost equally, and so we
divide these numbers in half to obtain the $t\anti t\gam\gam$
mode results of $S=130$ and $B=21$. To quantify our ability
to discriminate the SM case of $c=1,d=0$ from the 
$c=d=1/\sqrt 2$  and $c=0,d=1$ cases, we compute $D_1$ and $D_2$.
%assuming that the signal rate in the non-SM-like cases is the
%same as in the SM case.  
The resulting numerical values for $D_{1,2}$
are given in Table~\ref{disctrue} and indicate
the level of statistical significance at which an observed CP-even 
$\h$ could be said to {\it not} be (1) an equal mixture of CP-odd
and CP-even or (2) a purely CP-odd state.
%that accidentally yielded the same signal event rate.
These values may be somewhat optimistic in that the above rates
are those obtained before reconstructing the $t$ and $\anti t$.
However, because of the cleanliness of
the $t\anti t \gam\gam$ final state, we do not anticipate a large event
rate loss for such reconstruction.

As anticipated, the operators $a_1$, $b_1$, $b_2$ and $b_4$ provide
the best discrimination. For the best single operator, $b_1$,
discrimination from the $c=d=1/\sqrt 2$ case is only achieved at the 
$\sim 1.5\sigma$ level. To reach the more satisfactory $3\sigma$ level
would require of order four times as much integrated luminosity.
Discrimination between the purely CP-even case and the purely CP-odd case
would be possible using $b_1$ 
at a high level of statistical significance, $\sim 7\sigma$.
Better statistical significance in both cases can be achieved to
the extent that the different operators are sensitive to different aspects
of the $t$ and $\anti t$ distributions in the final state.  In particular,
$b_4$ is primarily sensitive to the $t,\anti t$ 
longitudinal momenta distributions, as distinct from  the
transverse momenta distributions that determine $b_1$. 
Simply combining the statistics for these two different moments
would imply that $c=1,d=0$ could be distinguished from $c=d=1/\sqrt 2$
at nearly the $2\sigma$ level, assuming the $L=600\fbi$ SM-like event rate.
In reality, the experimental groups would undoubtedly obtain the 
best level of discrimination by studying the likelihood that
a particular $c,d$ mixture fits their data in an overall sense
(without assuming knowledge of normalization).

A second possible channel for this type of analysis is the $X=b\anti b$
channel (\ie\ $t\anti t b\anti b$ final state). 
The signal event rate is much larger than in the $t\anti t\gam\gam$
final state, but there are
large backgrounds. This channel for Higgs discovery was first
explored in Refs.~\cite{DGV1,DGV2}, and has been studied by the ATLAS and
CMS collaborations with generally encouraging results.
These analyses isolate a signal by demanding that one of the $t$'s
decay semi-leptonically, and that three or four $b$-quarks
be tagged. The expected $b$-tagging efficiency and purity
at high luminosity are sufficiently large
that statistically significant signals for a SM-like $\h$ can be achieved.
In Ref.~\cite{ATLAS}, with $L=100\fbi$ and using 3-$b$-tagging, 
event rates are large enough that 
a roughly $5\sigma$ signal is seen for $\mh\sim 100\gev$, despite
a small $S/B\sim 1/40$ signal to background ratio. The 4-$b$-tagging
results of Ref.~\cite{DGV1} yield a cleaner $S/B\sim 1/2$, but
$S$ is also reduced so that $S/\sqrt B$ for $L=100\fbi$ is 
at most increased to $S/\sqrt B\sim 8-10$. (This latter study is not
a full experimental simulation, and it is quite possible that 
these larger $S/\sqrt B$ values are too optimistic.)
Neither of these analyses demands that the $t$-quarks be reconstructed,
as would be required in order to construct the $\ocp$ operators
considered here.  Such reconstruction could well improve the $S/B$
ratios (in particular, by largely eliminating the combinatoric
backgrounds), but undoubtedly at further sacrifice of signal event rate.  
Nonetheless, it seems probable that, for the combined
ATLAS+CMS benchmark luminosity of $L=600\fbi$, $S/\sqrt B\geq 5$ can be achieved
after full $t$-quark reconstruction. We assume that after such reconstruction
the only significant background 
will be from the irreducible $t\anti t b\anti b$ background process.
(This is clearly the case if 4-$b$-tagging is employed.)
In analyzing the experimental
error on the determination of $\asyms$ for the different $\ocp$
operators, we then need only compute the $\asymb$ and $\bsymb$
values for the irreducible $t\anti t b \anti b$ background.

\begin{table}[hbt]
\caption[fake]{$t\anti t b\anti b$ channel 
$\asymb$ and $\bsymb$ values, assuming $\mh=100\gev$.}
\begin{center}
\begin{tabular}{|c|cccccc|}
\hline
$\ocp$ & $a_1$ & $b_1$ & $b_2$ & $a_2$ & $b_3$ & $b_4$ \\
\hline
$\asymb$ & $-0.552$ & $-0.478$ & $-0.177$ & $-0.395$ & $-0.239$ & $0.241$ \\
$\bsymb$ & 1 & 0.642 & 0.122 & 1 & 0.284 & 0.375 \\
\hline
\end{tabular}
\end{center}
\label{abbkgndvaluesbb}
\end{table}

The values of $\asymb$ and $\bsymb$ for the cuts
$|y_{t,\anti t,b,\anti b}|<4$ and $p_{b,\anti b}^T>\mh/4$
are given in Table~\ref{abbkgndvaluesbb}.  
The corresponding signal reaction 
$\asyms$ and $\bsyms$ values for the various $\ocp$ 
operators are unchanged from the results appearing in
Table~\ref{absigvalues}.
Although the cuts
that will be required in the eventual analysis will be far more complex,
the changes in $\asyms$, $\bsyms$, 
$\asymb$ and $\bsymb$ are unlikely to be so large
as to invalidate the estimates obtained below.

Given $\asyms$, $\bsyms$, $\asymb$, and $\bsymb$, 
it is straightforward to compute the signal rate $S$ 
for a SM-like Higgs that would
be required to achieve $D_1=2$ (that is to discriminate
a pure CP-even coupling from an equal CP-even/CP-odd mixture at the $2\sigma$
statistical level) as a function of $B/S$. The results are easily summarized.
For $B/S\sim 1$, $D_1=2$ requires $S\sim 700$ ($\sim 600$),
equivalent to $S/\sqrt B\sim 26$ ($\sim 24$),
using $\ocp=b_1$ ($b_4$).
For $B/S\sim 50$, $D_1=2$ requires $S\sim 20,000$ ($\sim 15,000$),
equivalent to $S/\sqrt B\sim 20$ ($\sim 17$),
for $\ocp=b_1$ ($b_4$). The $S$ values for $b_4$ are a bit smaller
than for $b_1$, but probably not enough to compensate for 
inaccuracies associated with the need
to use longitudinal momenta in constructing the former. Note
the approximate rule that, for large $B/S$, $S/\sqrt B\sim 20$ is inevitably
required to achieve $D_1\sim 2$ (using $b_1$). 
The signal event rate required for $D_2=2$, \ie\ to distinguish
a purely CP-even Higgs from a purely CP-odd Higgs
at the $2\sigma$ level, is only about 1/20 
that needed to achieve $D_1=2$ (for a given $B/S$).  Indeed, $D_2\sim 2$
is achieved whenever we have a $S/\sqrt B\sim 4-5$ Higgs boson signal!

The above rules arise because $D_{1,2}$
scale as $S/\sqrt B$ for large $B/S$, see Eq.~\ref{dasymdef}.
Considering both the $\gam\gam$ and $b\anti b$
channels and both $D_1$ and $D_2$, we find, at large $B/S$,
\begin{equation}
\begin{array}{lll}
\gam\gam:& D_1=2 {S/\sqrt B \over 13}\,, & D_2=2{S/\sqrt B\over 2.7}\,\\
b\anti b:& D_1=2 {S/\sqrt B \over 22}\,, & D_2=2{S/\sqrt B\over 4.4}\,\\
\label{thumb}
\end{array}
\end{equation}
when employing the operator $b_1$ (alone).
In fact, the $S/\sqrt B$ values required for a given $D_{1,2}$
value are remarkably independent
of $B/S$ once $B/S\gsim 2$. For example, in the $t\anti t b\anti b$
channel, if $B/S\sim 2$ then $D_2=2$ is achieved for $S/\sqrt B\sim 5$.
Of course, these large $B/S$ scaling laws are not likely to be relevant for
the $t\anti t\gam\gam$ channel, since $B/S\ll 1$ is the general rule there.
Once $B/S\ll 1$, $D_{1,2}$ scale as $\sqrt S$:
in the limit of $B=0$, one
simply uses Table~\ref{discpower} to obtain the $S$ required
for a given $D_{1,2}$.

Overall, the $t\anti t b\anti b$ channel will probably
not be as useful as the $t\anti t \gam\gam$ channel in the case of
a SM-like CP-even Higgs boson. However, for a non-SM-like Higgs boson,
$BR(\h\to \gam\gam)$ might be too small to yield an observable signal 
in the $t\anti t\gam\gam$ final state, 
whereas $BR(\h\to b\anti b)$ will generally be large
and an observable signal in the $t\anti t b\anti b$ channel will
be possible so long as the $t\anti t\h$ coupling is not significantly
suppressed compared to SM strength.  The most difficult example is the case
of a purely CP-odd $\ha$. The absence of the $\wp\wm\ha$ coupling
implies very small $BR(\ha\to\gam\gam)$.  In a two-Higgs-doublet
model of type-II, the $t\anti t\ha$ ($b\anti b\ha$) 
coupling is $\propto \cot\beta$ ($\propto\tan\beta$),
where $\tanb$ is the ratio of the vacuum expectation values of the two
neutral Higgs doublet fields \cite{hhg,dpfreport}.  
In the preferred $\tanb>1$ region of parameter space, 
the $t\anti t\ha$ production cross section, proportional to
$\cot^2\beta$, is suppressed, but $BR(\ha\to b\anti b)$
is large (for $\mha$ below $2\mt$). With $L=600\fbi$, it might prove
possible to achieve $S/\sqrt B\sim 5$ for $\tanb\lsim 2$, 
which (assuming large $B/S$ and
employing $\delta\asyms(c=0,d=1)$ for $b_1$ from 
Table~\ref{absigvalues} in defining $D_2$) would imply $D_2\sim 1.6$.
This would at least add support to the indirect 
evidence for a large CP-odd component
deriving simply from the absence of observable $W+$Higgs and 
$t\anti t \gam\gam$ channel signals.

\section{Final Remarks and Conclusions}

\indent\indent
In this letter, we have shown that weighted moments of the $t\anti t\h$
production cross section can provide a rough but direct
(cross section normalization independent) determination of the CP
nature of a light Higgs boson at the LHC.
In the $t\anti t\gam\gam$ final state, if we employ
the best single weighting operator
that does not depend upon the longitudinal momenta of the $t$ quarks,
then a SM-like Higgs boson
(pure CP-even) can be distinguished from a Higgs boson
that is an equal mixture of CP-odd and CP-even at the $\sim 1.5\sigma$
statistical level with about 130 signal events (assuming $B/S\ll 1$).
This is roughly the number obtained for a combined ATLAS+CMS luminosity of
$L=600\fbi$ (assuming that the required $t$-quark reconstruction
does not cause large losses in this final state).
About 240 signal events (after $t$-quark reconstruction)
are needed to achieve $2\sigma$ discrimination.
These same numbers of events would distinguish
the purely CP-even Higgs from a purely CP-odd Higgs at the 
$\sim 7\sigma$ and $\sim 10\sigma$ level, respectively. If the
best operator depending upon $t$-quark longitudinal momenta can 
also be employed, the 
above statistical significances would increase by about 40\%.

In the $t\anti t b\anti b$ channel, typically characterized
by $B/S\gsim 1$ (possibly $\gg 1$), $2\sigma$ discrimination between
a pure CP-even Higgs and 
one with an equal mixture of CP-odd and CP-even components requires
$S/\sqrt B\sim 20$, which would in general be very difficult
to achieve (except for a Higgs with enhanced $t\anti t \ha$
coupling --- requiring $\tanb<1$ in the CP-violating
type-II two-Higgs-doublet model).
However, distinguishing between purely CP-even and purely CP-odd
coupling at the $\gsim 2\sigma$ level is possible whenever $S/\sqrt B\gsim 4$,
that is, whenever the Higgs boson can be detected. 

If both the $t\anti t \gam\gam$ and $t\anti t b\anti b$ channels yield
a useful level of discrimination, statistics in the two channels
can be combined to further improve the overall discrimination level.

Although our analysis has made use of rather simplified cuts, we do not 
believe that these results depend very much on the cuts.  Nonetheless,
the experimental groups should compute the weightings
defined in Eqs.~\ref{adef} and \ref{bdef},
for both the $t\anti t\gam\gam$ and $t\anti t b\anti b$ channels,
using their full simulation, including the necessary $t$-quark
reconstruction.

A more detailed discussion of the implications of these results for 
a general CP-violating two-Higgs-doublet
model will appear in a later paper \cite{later}.

\section{Acknowledgements}

This work was supported in part by Department of Energy under
grant No. DE-FG03-91ER40674
and by the Davis Institute for High Energy Physics. 
XGH was supported in part by the Australian Research Council. XGH
would like to thank the Davis Institute for High Energy Physics for
hospitality.

\clearpage
 
\end{document}